%% file: main.tex
\documentclass[conference,dvipsnames]{IEEEtran}

\usepackage{booktabs} 
\usepackage{makecell}
\usepackage{listings}
\usepackage{courier}
\usepackage{balance}
\usepackage[normalem]{ulem}
\usepackage[ruled,noline,noend,linesnumbered]{algorithm2e}
\usepackage{multirow}
\usepackage{multicol}
\usepackage{graphicx}
\usepackage{subcaption}
\usepackage{soul}
\usepackage{amsmath}
\usepackage{comment}
\usepackage{color}
\usepackage[frozencache=true,cachedir=_minted-main]{minted}

\newcommand\sd[1]{{\color{blue}#1}} 
\definecolor{pinegreen}{rgb}{0.0, 0.47, 0.44}
\newcommand{\code}[1]{\texttt{#1}}

\begin{document}

\title{A Unified System for Data Analytics and In Situ Query Processing 
}


\author{\IEEEauthorblockN{Alex Watson\textsuperscript{\textsection}, Suvam Kumar Das\textsuperscript{\textsection}, and Suprio Ray} 
\IEEEauthorblockA{\textit{Faculty of Computer Science, University of New Brunswick, Canada.} 
Email: \{awatson, suvam.das, sray\}@unb.ca}
}

\maketitle

\begingroup\renewcommand\thefootnote{\textsection}
\footnotetext{Equal contribution}
\endgroup

\begin{abstract}
In today’s world data is being generated at a high rate due to which it has become inevitable to analyze and quickly get results from this data. Most of the relational databases primarily support SQL querying with a limited support for complex data analysis. Due to this reason, data scientists have no other option, but to use a different system for complex data analysis. Due to this, data science frameworks are in huge demand. But to use such a framework, all the data needs to be loaded into it. This requires significant data movement across multiple systems, which can be expensive.

We believe that it has become the need of the hour to come up with a single system which can perform both data analysis tasks and SQL querying. This will save the data scientists from the expensive data transfer operation across systems. In our work, we present DaskDB, a system built over the Python’s Dask framework, which is a scalable data science system having support for both data analytics and in situ SQL query processing over heterogeneous data sources. DaskDB supports invoking any Python APIs as User-Defined Functions (UDF) over SQL queries. So, it can be easily integrated with most existing Python data science applications, without modifying the existing code. Since joining two relations is a very vital but expensive operation, so a novel distributed learned index is also introduced to improve the join performance. Our experimental evaluation demonstrates that DaskDB significantly outperforms existing systems.
\end{abstract}

\input{intro}

\input{background}
\input{related}
\input{approach}

\input{evaluation}

\input{conclusion}

\bibliographystyle{IEEEtran}
\bibliography{bibliography} 

\end{document}

%% file: intro.tex
\vspace{-0.1cm}
\section{Introduction}
\label{sec:introduction} 


Due to the increasing level of digitalization, large volumes of data are constantly being generated.  To make sense of the deluge of data, it must be cleaned, transformed and analyzed.  Data science offers tools and  techniques to manipulate data in order to extract actionable insights  from data.
These include support for data wrangling,  statistical analysis and machine learning model building.  
Traditionally, practitioners and researchers 
make a distinction between  query processing and data analysis tasks. Whereas relational database systems (henceforth, databases or DBMSs) are used for SQL-style query processing, a separate category of frameworks are used for data analyses  that include statistical and machine learning tasks. Currently, Python has emerged as the most popular
language-based framework, for its rich ecosystem of data analysis libraries, such as Pandas, Numpy, scikit-learn. These tools make it possible to perform \textit{in situ}  analysis of  data that is stored outside of any database.
However, a significant amount of data is still stored in databases. To do analysis on this data, it must be moved from a database into the address space of the data analysis application.
Similarly, to do SQL query processing on data that is stored in a raw file, it must be loaded into a database using a  loading mechanism, which is  known as ETL (extract, transform, load). This movement 
 of data and loading of data are both time consuming operations. 

To address the movement of data across databases and data analysis frameworks, recently researchers have proposed several approaches. Among them, a few are in-database solutions, that incorporate data analysis functionalities within an existing database. These include PostgreSQL/Madlib~\cite{MadSkillsHellerstein2012},  HyPer~\cite{Hyper2011} and AIDA~\cite{D'silva:2018:AIDA}. In these systems, the application developers write SQL code and invoke data analysis functionalities through user-defined functions (UDF). There are several issues with these approaches. \textbf{First}, the vast body of existing data science applications that are  written in a popular language (Python or R), need to be converted into SQL. \textbf{Second}, the data analysis features supported by databases are usually through UDF functions, which are not as rich as that of the language-based API ecosystem, such as in Python or R. \textbf{Third}, data stored in raw files needs to be loaded into a database through ETL. Although, some support for executing SQL queries on raw files exist, such as PostgreSQL's support for foreign data wrapper, this can easily break if the file is not well-formatted. 
In recent years several projects~\cite{AlagiannisBBIA12NoDB, 
Cheng2014Parallel, OlmaKAAA20Adaptive} investigated how to support \textit{in situ} SQL querying on raw data files. However, they primarily focused on supporting database-like query processing,  operating on a single machine.
These systems lack sophisticated data wrangling and data science features that is available in Python or R.
 \textbf{Fourth}, most  relational databases are not horizontally scalable. Even with parallel databases, the  parallel execution of UDFs is either not supported or not efficient. Although ``Big Data'' systems such as Spark \cite{SparkZaharia2010} and 
  Hive/Hivemall~\cite{url:hivemall}
 address some of these issues,
 that they often involve more complex APIs and  a  steeper  learning  curve. Also, it is not practical to  rewrite the large body of existing data science code with these APIs.

To address the issues with the existing approaches, we introduce a scalable data science system, DaskDB, which seamlessly supports \textit{in situ} SQL query execution and data analysis using Python. DaskDB extends the scalable data analytics framework Dask~\cite{Rocklin2015Dask} 
 that can scale to more than one machine. Dask's high-level collections APIs mimic many of the popular Python data analytics library APIs based on Pandas and NumPy. So, existing applications written using Pandas collections need not be modified. 
On the other hand, Dask does not support SQL query processing. In contrast, 
DaskDB can execute SQL queries \textit{in situ}  without requiring the expensive ETL step and  movement of data from raw files into a database system.  Furthermore, with DaskDB, SQL queries can have UDFs that directly invoke Python 
APIs. This provides a powerful mechanism of mixing SQL with Python and enables data scientists to take advantage of the rich data science libraries with the convenience of SQL. Thus, DaskDB unifies query processing and  analytics in a scalable manner.

A key issue with distributed query processing and data analytics is the movement of data across nodes, which can significantly impact
the performance. 
We propose a \textit{distributed learned index}, 
to improve the performance of join that is an important data operation.
 In DaskDB, a relation (or dataframe) is split into multiple partitions, where each partition consists of numerous tuples of that relation. These partitions are distributed across different nodes. 
 While processing a join, it is possible that not all partitions of a relation contribute to the final result when two relations are joined. 
 The \textit{distributed learned index} is 
 designed to efficiently consider only those partitions that contain the required data in \textit{constant} time, by identifying the data pattern  in each partition. This minimizes the unnecessary data movement across nodes.
DaskDB also incorporates selective data persistence that significantly reduces serialization/de-serialization overhead and data movement.

We conduct extensive experimental evaluation to compare the performance of DaskDB against two horizontally scalable systems: PySpark and Hive/Hivemall. Our experiments involve workloads from a few queries from  TPC-H~\cite{url:TPCH} benchmark, with different data sizes (scale factors). We also created a custom UDF benchmark to evaluate DaskDB and PySpark. Our results show that DaskDB outperforms others in both of these benchmarks. 
%
The key contributions of this paper are: 
\vspace{-2pt}
\begin{itemize}
 \setlength{\itemsep}{1pt}
  \setlength{\parskip}{0pt}
  \setlength{\parsep}{0pt}
  
\item We propose DaskDB that integrates \textit{in situ} query processing and data analytics in a scalable  manner.
\item DaskDB supports SQL queries with UDFs that can directly invoke Python data science APIs.
\item We introduce a novel distributed learned index.
\item We present extensive experimental results involving TPC-H benchmark and a custom UDF benchmark.
\end{itemize}


%% file: background.tex
\section{Background}
\label{sec:background} 

DaskDB was built by extending the Dask \cite{Rocklin2015Dask}, which is an open-source library for parallel computing in Python. 
The main advantage of Dask is that it provides Python APIs and data structures that are similar to Numpy, Pandas, and Scikit-Learn. Hence, programs written using Python data science APIs can easily be switched to Dask by changing the import statement. Dask comes with an efficient task scheduler, which can run programs on a single node  and scale to many nodes.

However, Dask does not support SQL queries. 
We show that DaskDB, which is 
built over Dask, can outperform Spark~\cite{url:Spark}.

%% file: related.tex
\section{Related Work}
\label{sec:related} 


In this section, 
first we discuss about systems to perform data analytics and query processing. Next, we look at works related to learned index, followed by in situ query processing. 

\subsection{Data Analytics and Query Processing}
\label{related_data_analytic_query_processing}


\subsubsection{Dedicated Data Analytics Frameworks}
\label{related_dedicated_analytic_systems}

Many open-source data analytic applications traditionally use R.
More recently, Python has become very popular because of the Anaconda distribution \cite{url:anacondaPython}. 
It contains many data science and analytics packages,
such as Pandas, SciPy, Matplotlib, and scikit-learn.
Some popular commercial data analytic systems include Tableau \cite{url:tableau} and  MATLAB \cite{url:MATLAB}.

\subsubsection{In-Database Analytics}
\label{related_in_database_ana}
An increasing number of the major  DBMSs now include data science and machine learning tools. 
For instance, PostgreSQL supports SQL-based algorithms for machine learning
with the Apache MADlib library \cite{MadSkillsHellerstein2012}.
However, interacting with a DBMS to implement analytics can be challenging \cite{Fouch2018Indatabaseibmdbpy}. 
Although SQL is a mature technology, it is not rich enough  for extensive data analysis.
DBMSs typically  support 
analytics functionalities through User Defined Functions (UDF). Since, a UDF may execute any external code written in R, Python,  Java or T-SQL, a DBMS treats a UDF as a black box because no optimization can be performed on it.
It is also difficult  to debug  and  to  incrementally  develop~\cite{D'silva:2018:AIDA}.
%
The in-database analytics approaches still have the constraint of ETL (extract, transform, load), which
is a time-consuming process and not practical in many cases. 


\subsubsection{Integrating Analytics and Query Processing}
\label{sec:rel_integrating}
There have been  several attempts at creating more efficient  solutions and they  combine two or more of either dedicated data analytic systems, DBMS or big data frameworks.  These systems can be  classified into 2 categories that we describe next.

\textbf{Hybrid Solutions.} These solutions integrate two or more system types together into one and are primarily DBMS-centric approaches. AIDA \cite{D'silva:2018:AIDA} integrates a Python client directly to use the DBMS memory space, eliminating the bottleneck of transferring data.
In \cite{Lajus:2014RStudioMonetDB} the authors present a prototype system that integrates a columnar relational database (MonetDB) and R together using a  same-process zero-copy data sharing mechanism. 
In \cite{Raasveldt2020DataMF}, the authors proposed an embeddable analytical database DuckDB.
The key drawback of these hybrid systems is ETL, since the data needs to be loaded into a database. Moreover, existing data science applications written in Python or R, need to be modified to work in such systems, since their   interface is SQL-based. 


\textbf{``Big Data'' Analytics Frameworks.}
 The most popular big data frameworks are Hadoop 
 and Spark \cite{SparkZaharia2010}. Spark supports machine learning with MLlib \cite{SparkMLlib2016} and
 SQL like queries. 
  Hive is based on Hadoop that supports SQL-like queries and supports analytics with the machine learning library Hivemall~\cite{url:hivemall}.
 Some drawbacks of big data frameworks include more complicated development and steeper learning curve than most other analytics systems and the difficulty in integration with DBMS applications.  
To run any existing Python or R application within a  big data system, it will require rewriting these applications with new APIs. This is not a viable option in most cases.


\subsection{Learned Index}
\label{sec:rel_learned_index}
Data structures such as B+trees are the mainstay of indexing techniques. These approaches require the storage of all keys for a dataset.
Recent studies have shown that 
learned models can be used to model the cumulative distribution function (CDF) of the keys in a sorted array. This can be used to predict their locations for the purpose of indexing and this idea was termed as learned index~\cite{LearnedIndexStructures}.
Several learned indexes  were developed that include FITing-Tree \cite{FittingTree} and PGM-Index \cite{PGM-Index}.

The learned index approaches proposed so far were meant only for stand-alone systems.
These ideas have not been incorporated as part of any database system yet, to the best of our knowledge. Also, no learned index model has yet been developed for any distributed data system.

\subsection{In Situ Query Processing}
\label{sec:rel_in-situ_query}


A vast amount of data is stored in raw file-formats that are not inside traditional databases. Data scientists, who frequently lack expertise in data modeling, database admin and ETL tools,  often need to run interactive analysis on this data. To reduce the ``time to query'' and avoid the overhead associated with relational databases, a number of research projects investigated \textit{in situ} query processing on raw data.  

NoDB~\cite{AlagiannisBBIA12NoDB} was  one of the earliest systems to support in situ query processing on raw data files. 
PostgresRaw~\cite{url:PostgresRAW} is based on the idea of NoDB and it supports  SQL querying over CSV files in PostgreSQL. 
The SCANRAW~\cite{Cheng2014Parallel}  system exploits parallelism during in situ raw data processing. Slalom~\cite{OlmaKAAA20Adaptive} introduced adaptive partitioning and on-the-fly per partition indexes to improve query processing performance on raw data. All these systems were  focused on database-style SQL query processing on raw data and on a single machine. Our system, DaskDB supports \textit{in situ} querying on heterogeneous data sources, and it also supports doing data science.
Moreover, it is a distributed data system that can scale over a node cluster.

%% file: approach.tex
\section{Our Approach: DaskDB}

In this section, we present DaskDB. 
DaskDB, in addition to supporting all Dask features, also enables \textit{in situ} SQL querying on raw data in a data science friendly environment. 
Next, we describe DaskDB system architecture and its components.

\begin{figure}[t!]
\centering
\includegraphics[width=0.7\linewidth]{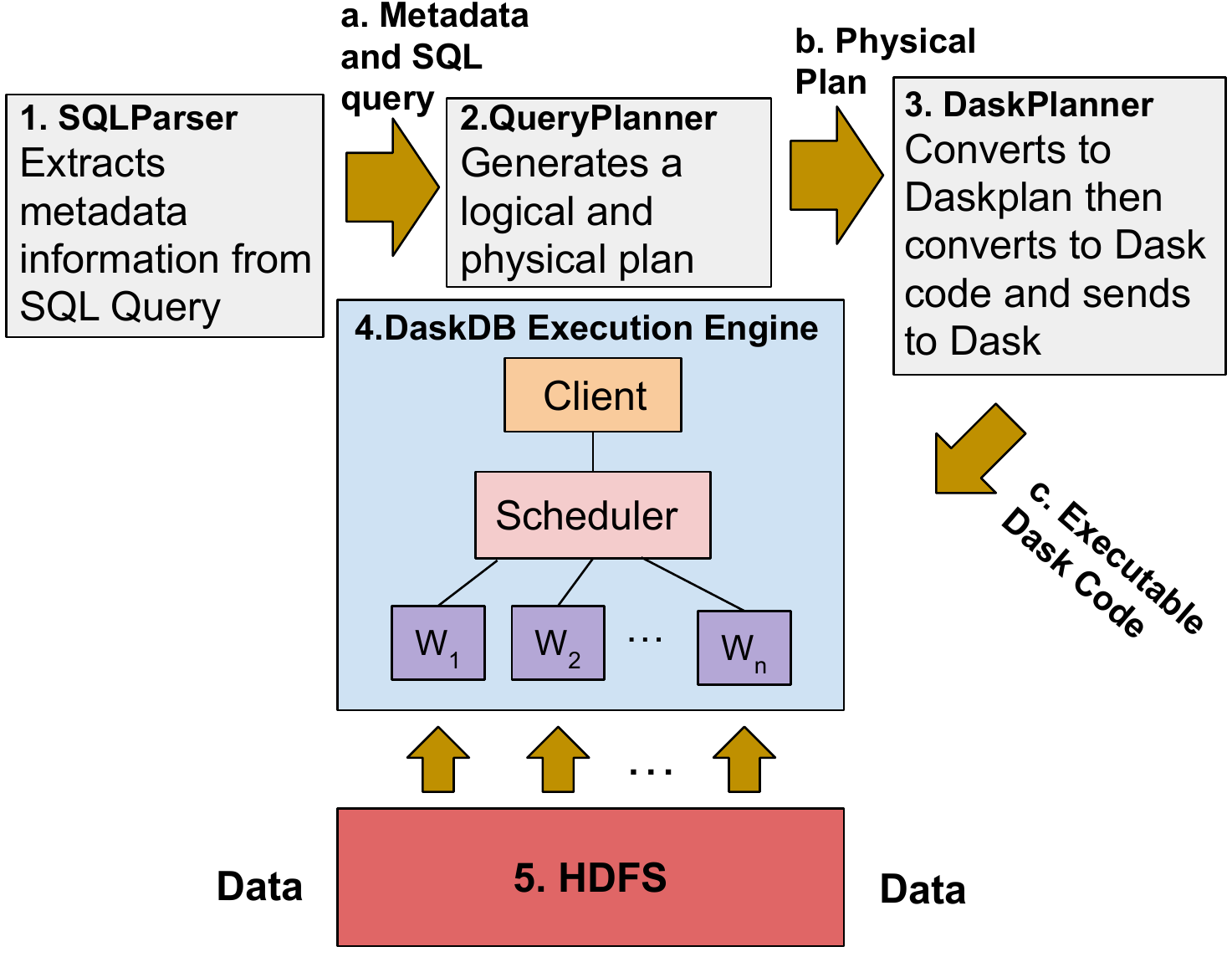}
\caption{DaskDB System Architecture}
\label{fig:daskdb_architecture}
\vspace{-10pt}
\end{figure}

\subsection{System Architecture}
\label{sec:system_arch}

The system architecture of DaskDB incorporates five main components: the SQLParser, QueryPlanner, DaskPlanner, DaskDB Execution Engine, and HDFS. They are shown in Figure \ref{fig:daskdb_architecture}.
First, the SQLParser gathers metadata information pertaining to the SQL query, such as the names of the tables, columns and functions. This information is then passed along to the QueryPlanner. Next, in the QueryPlanner component, physical plan is generated from the information sent by SQLParser about the SQL query. The physical plan is an ordered set of steps that specify a particular execution plan for a query and how data would be accessed. The QueryPlanner then sends the physical plan to the DaskPlanner. In the DaskPlanner, a plan is generated, which includes operations that closely resemble Dask APIs, called the \textit{Daskplan}. The Daskplan is produced from the physical plan, and it is then converted into Python code and sent to DaskDB Execution Engine.
DaskDB Execution Engine then executes the code and gathers the data from the HDFS, and thus executes the SQL query. Further details are provided in the next sections.

\subsubsection{SQLParser}
The SQLParser is the first component of DaskDB that is involved in query processing.  The input for the SQLParser is the original SQL query. It first checks for syntax errors and creates a parse tree with all the metadata information about the query. We then process the parse tree to gather the metadata information needed by the QueryPlanner. This metadata information includes table names, column names and UDFs. We then check if the table(s) exist 
and if they do, we 
dynamically generate a schema. The schema contains information about  tables and column names and data types used in the SQL query.  The schema, UDFs (if any) and the original SQL query are then passed to the QueryPlanner.
 

\subsubsection{QueryPlanner}

The QueryPlanner
creates logical and preliminary physical plans. The schema and UDFs produced by SQLParser, along with the SQL query, are passed into the QueryPlanner. The QueryPlanner uses these to first create a logical plan and then an optimized preliminary physical plan. This plan is then sent to the DaskPlanner.


\subsubsection{DaskPlanner}

\begin{algorithm}
  \KwIn{A physical plan (\textit{P}) is given as input. \textit{P} contains ordered groups of dependent operators (\textit{G}). Each \textit{G} consists of an ordered list of tuples (\textit{k, o, d}), where \textit{k} contains the unique key of a operation, \textit{o} is the operation type and \textit{d} contains the operation metadata information. }
  \KwOut{The final result is a Daskplan \textit{DP}, which consists of an ordered list of operators.}
  \DontPrintSemicolon
  DP $\leftarrow$ list()\;
  \For{sorted($G \in P$)}{ 
    \For{sorted($k, o,d \in G$)}{ 
        dp $\leftarrow$ dict() //create Daskplan operator\;
        dp[o] $\leftarrow$ convertToDaskPlanOperator(o) \;
        dp[d] $\leftarrow$ getMetadataInfo(d) \;
        dp[key] $\leftarrow$ k //adds key (used to get data dependencies)  \;
        DP.add(dp) //adds dp to Daskplan  \;
    }
  }
  
  \For{sorted($dp \in DP$)}{ 
    \While{$dp$ has children (c)}{
        dp[t$_i$] $\leftarrow$ get table information from c$_i$\;
    }
  }
  return \textbf{DP}\;
  \caption{DaskPlanner: Conversion of Physical Plan to Daskplan}
  \label{algo1:phystoDask}
\end{algorithm}

The DaskPlanner is used to transform the preliminary physical query plan from the QueryPlanner into Python code that is ready for execution. The first step in this process is for the DaskPlanner to go through the physical plan obtained from QueryPlanner and convert it into 
a Daskplan. This maps the operators from the physical plan into operators that more closely resemble the Dask API.  This Daskplan also associates relevant information with each operator from the physical plan. This information includes columns and tables involved and specific metadata information for a particular operator. We also keep track of each operator's data dependency. This is needed to pass intermediate results from one operation to the next.  Algorithm \ref{algo1:phystoDask} shows how DaskDB converts the physical plan into the Daskplan.

In the next step, the DaskPlanner converts the Daskplan into the Python code, which utilizes the Dask API. 
All of the detail about each table and their particular column names and indexes are tracked and maintained in a dynamic dictionary throughout the execution of 
a query. This is because multiple tables and columns may be created, removed or manipulated during execution. 
For these reasons, the names of the tables, columns, and indexes are dynamically maintained while transforming the Daskplan into Python code to execute it.

\subsubsection{DaskDB Execution Engine}
There are three main components of the DaskDB execution engine: the client, scheduler and workers. The client transforms the Dask Python code into a set of tasks. The scheduler creates a DAG (directed acyclic graph) from the set of tasks, automatically partitions the data into chunks, while taking into account data dependencies. The scheduler sends a task at a time to each of the workers according to several scheduling policies. The scheduling policies
for task and workers depend on various factors including data locality. 
A worker 
stores a data chunk until it is not needed anymore and is instructed by the scheduler to release it. 

 

\subsubsection{HDFS}

The Hadoop Distributed File System (HDFS)~\cite{ShvachkoHadoop2010} 
is a  storage system used by Hadoop applications. 
DaskDB uses HDFS to store and share the data files among its nodes. 

\subsection{Illustration of SQL query execution} 
\label{sec:qry_illustration}

An \textit{in situ} query is executed within DaskDB by calling \code{query} function with the SQL string as argument.  The query in  Figure~\ref{Code:sql} is a simplified version of 
a typical TPC-H  query.

\begin{figure}[ht]
\begin{minted}
[fontsize=\scriptsize] %\footnotesize tiny
{python}
from daskdb_core import query

sql = """SELECT l_orderkey, sum(l_extendedprice * 
    (1-l_discount)) as revenue
  FROM orders, lineitem
  WHERE l_orderkey = o_orderkey and o_orderdate >= '1995-01-01'
  GROUP BY l_orderkey
  ORDER BY revenue LIMIT 5 ; """
query(sql)
\end{minted}
\caption{Code showing SQL query execution}
\label{Code:sql}
\end{figure}

\begin{figure}[t!]
\centering
\includegraphics[width=0.75\linewidth]{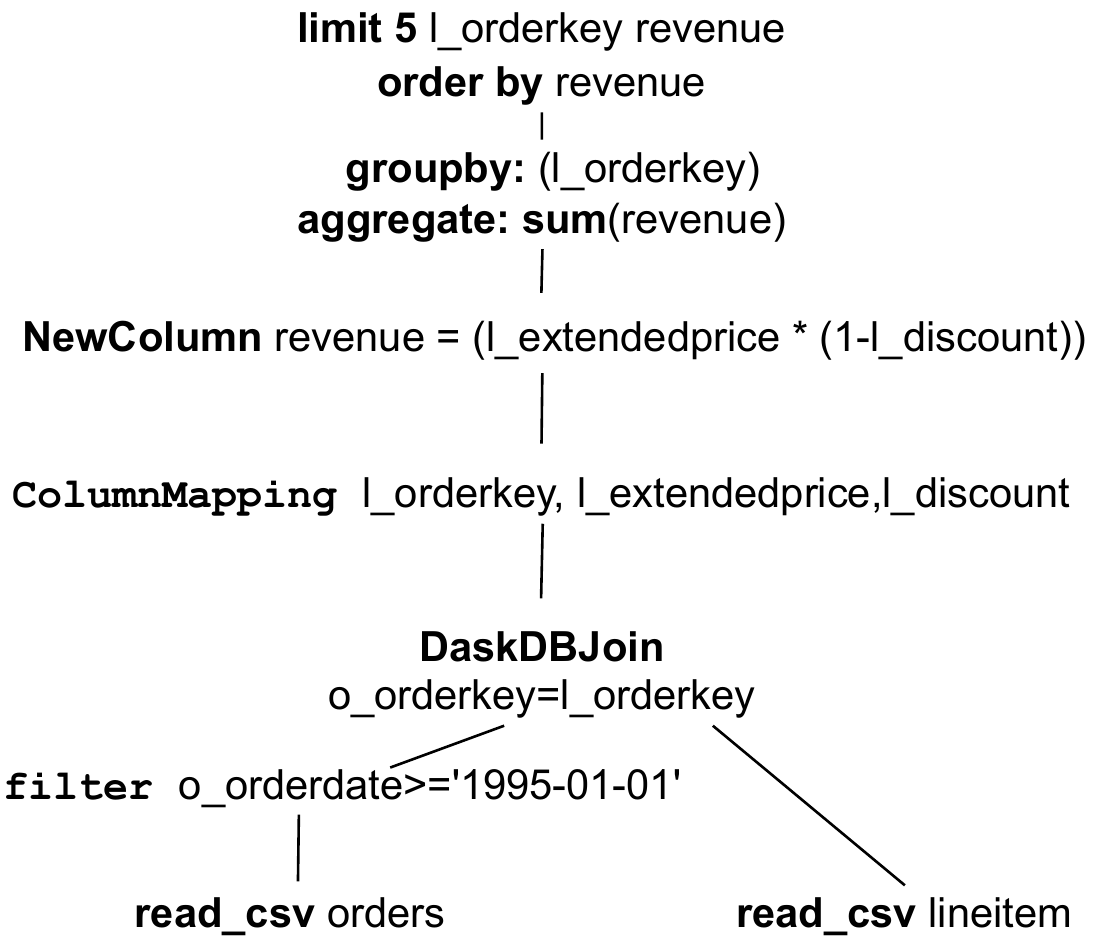}
\caption{Generated Daskplan for the code in Figure \ref{Code:sql}}
\label{subfig:DaskPlan}
\vspace{-10pt}
\end{figure}

The Daskplan, shown in Figure \ref{subfig:DaskPlan}, is generated from the physical plan in the DaskPlanner component. The Daskplan  operators more closely resemble the Dask API. For example, these include the \code{read\_csv} and \code{filters} methods shown in the tree. This Daskplan is then converted into executable Python code, which is omitted due to space constraint.

\subsection{Support for SQL query with UDFs}
DaskDB supports  UDFs in SQL as part of \textit{in situ} querying. A UDF enables a user to create a function using Python code and embed it into the SQL query. Since DaskDB converts the SQL query, including the UDF back into Python code, the UDFs can reference and utilize features from any of the data science packages available in Anaconda Python API ecosystem. 
Spark introduced UDF's since version 0.7 but it operated one-row-at-a-time and thus suffered from high serialization and invocation overhead. Later they came up with 
\textit{Pandas UDF} which provides low-overhead, high-performance UDFs entirely in Python. Although Pandas UDFs are efficient, but are restrictive to use in queries. Sometimes it also requires to use Spark's own data types, which would be inconvenient for users who are not experienced in Spark. In contrast, in DaskDB UDFs for SQL queries can easily be written. Any native Python function (either imported from an existing package or custom-made), which accepts Pandas dataframes as parameters can be applied as UDFs to the SQL queries in DaskDB. The return type of the UDFs is not fixed like Spark's Pandas UDF, and hence allows the user to design UDFs with ease. Like a general Python function, UDFs with code involving machine learning, data visualization and numerous other functionalities can easily be developed and applied on queries in DaskDB. 

\subsection{Illustration of SQL query with UDF} 
\label{sec:UDF_illustration}

We illustrate UDF with SQL in DaskDB with K-Means clustering. Similar to Spark, a UDF needs to be registered to DaskDB system using the \code{register$\_$udf} API. As shown in Figure \ref{Code:UDF_K-Means}, the UDF \code{myKMeans} takes as input a single Pandas dataframe having 2 columns; hence the UDF is registered as \code{register$\_$udf (myKMeans, [2])}. 
As part of the query, the UDF is invoked as \code{myKMeans(l$\_$discount, l$\_$tax)}, which means after application of the selection condition \code{(l$\_$orderkey $<$ 50)} and the limit \code{(limit 50)} to the \code{lineitem} relation, both the columns \code{l$\_$discount} and \code{l$\_$tax} together form a Pandas dataframe and is passed to \code{myKMeans}. The output of the code is plotted in Figure \ref{subfig:K-Means}, where the different clusters are shown by different colours.

\begin{figure}[ht]
\begin{minted}
[fontsize=\scriptsize] %\footnotesize
{python}
from daskdb_core import query, register_udf
import matplotlib.pyplot as plt
from sklearn.cluster import KMeans

def myKMeans(df):
    kmeans = KMeans(n_clusters=4).fit(df)
    col1 = list(df.columns)[0]
    col2 = list(df.columns)[1]
    plt.scatter(df[col1], df[col2],
        c= kmeans.labels_.astype(float), s=50)
    plt.xlabel(col1)
    plt.ylabel(col2)
    plt.show()
    
register_udf(myKMeans,[2])    
sql_kmeans = """select myKMeans(l_discount, l_tax)
 from lineitem where l_orderkey < 50 limit 50; """
query(sql_kmeans)
\end{minted}
\caption{UDF code showing K-Means Clustering}
\label{Code:UDF_K-Means}
\vspace{-8pt}
\end{figure}

\begin{table*}[ht]
\begin{minipage}[b]{0.6\textwidth}
\centering
\begin{tabular} { |p{0.12\textwidth}|p{0.8\textwidth}|}
    \hline
    \textbf{Tasks} & \textbf{Query} \\ \hline
    LR & select myLinearFit(l$\_$discount, l$\_$tax) from lineitem where l$\_$orderkey $<$ 10 limit 50 \\ \hline
    K-Means & select myKMeans(l$\_$discount, l$\_$tax) from lineitem, orders where l$\_$orderkey = o$\_$orderkey limit 50 \\ \hline
    Quantiles & select myQuantile(l$\_$discount) from lineitem, orders where l$\_$orderkey = o$\_$orderkey limit 50 \\ \hline
    CGO & select myCGO(l$\_$discount, l$\_$tax) from lineitem where l$\_$orderkey $<$ 10 limit 1\\ \hline
   \end{tabular}
    \caption{Queries with UDF}
    \label{table:queries_with_UDF}
\end{minipage}
 \hspace{25pt}%
\begin{minipage}[b]{0.3\textwidth}
\centering
    \includegraphics[width=50mm]{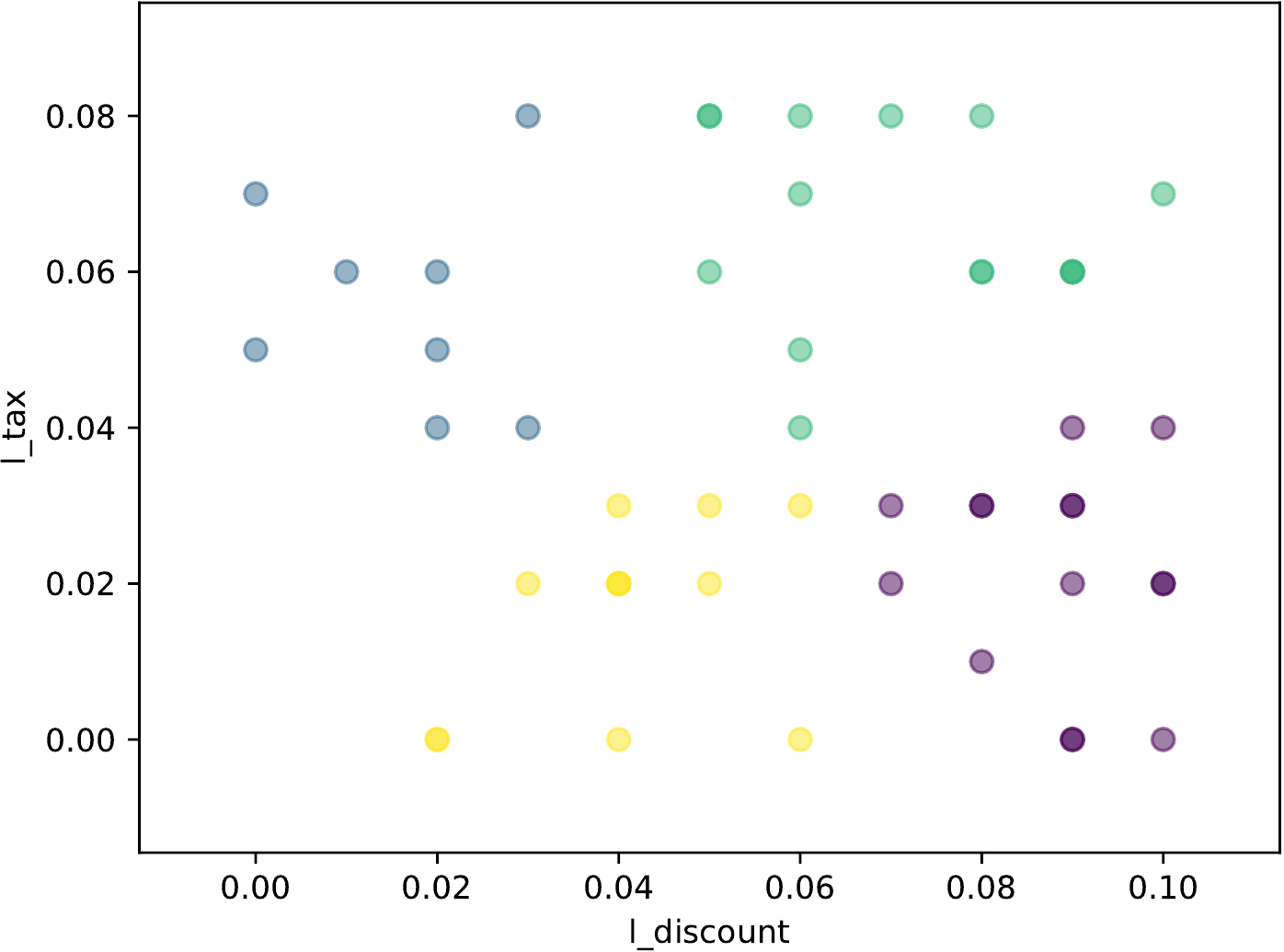}
    \captionof{figure}{K-Means output}
    \label{subfig:K-Means}
\end{minipage}
\vspace{-2pt} 
\end{table*}

\subsection{Distributed Learned Index}
\label{sec:distributed_learned_index}
 Join is considered one of the most expensive data operations.  We propose a novel distributed learned index, to accelerate distributed execution of join in DaskDB.
 %
Our distributed learned index relies on Heaviside step function~\cite{heaviside-function-url}. Any step function can be represented by a combination of multiple Heaviside step functions, which forms the basis of our learned index structure.
A Heaviside step function can be defined as 
\begin{equation}
H(x) = \begin{cases} 
      0 & x < 0 \\
      1 & x \geq 0 \\
   \end{cases}
 \end{equation}

\noindent We define a  \textit{Partition Function, P} as
\begin{equation} \label{dist_heaviside}
P_{a,b,c}(y) = H((b-y) * (y-a)) * c
\end{equation}

\noindent which returns \textit{c} whenever \textit{a $\le$ y $\le$ b} , or returns 0 otherwise.   

While constructing the distributed learned index it is assumed that one of the relations is sorted by the join attribute. Hence, if an index can maintain the first and last values of the keys for each partition, then given any key, the partition containing the key can be identified by a \textit{Partition Function}. A sparse index in this case can entail huge  storage savings, since all the keys are not required to be stored.
We illustrate this using a simplified example with the \textbf{customer} table from TPC-H, where \textit{c\_custkey} is the primary key. If there are 500 tuples in this relation and each table partition can store 100 tuples, then there will be total 5 partitions. The distribution of the keys is shown in Table \ref{table:sparse_index_unique_keys}, and also plotted in Figure \ref{fig:step_function}.

\begin{table}[ht]
\begin{minipage}[b]{0.4\linewidth}
\centering
    \begin{tabular}{rlrrrrrr}
    \hline
    \multicolumn{1}{c}{\textbf{Begin}} & \multicolumn{1}{c}{\textbf{End}} & \multicolumn{1}{c}{\textbf{Partition}}\\ \hline \hline

    \multicolumn{1}{c}{1} & \multicolumn{1}{c}{200} & \multicolumn{1}{c}{1} \\
    \multicolumn{1}{c}{250} & \multicolumn{1}{c}{380} & \multicolumn{1}{c}{2} \\
    \multicolumn{1}{c}{400} & \multicolumn{1}{c}{560} & \multicolumn{1}{c}{3} \\
    \multicolumn{1}{c}{580} & \multicolumn{1}{c}{700} & \multicolumn{1}{c}{4} \\
    \multicolumn{1}{c}{701} & \multicolumn{1}{c}{800} & \multicolumn{1}{c}{5} \\
   \hline
   \end{tabular}
    \caption{Sparse index for \textbf{customer} relation}
    \label{table:sparse_index_unique_keys}
\end{minipage}\hspace{8pt}
\begin{minipage}[b]{0.6\linewidth}
\centering
\includegraphics[width=45mm]{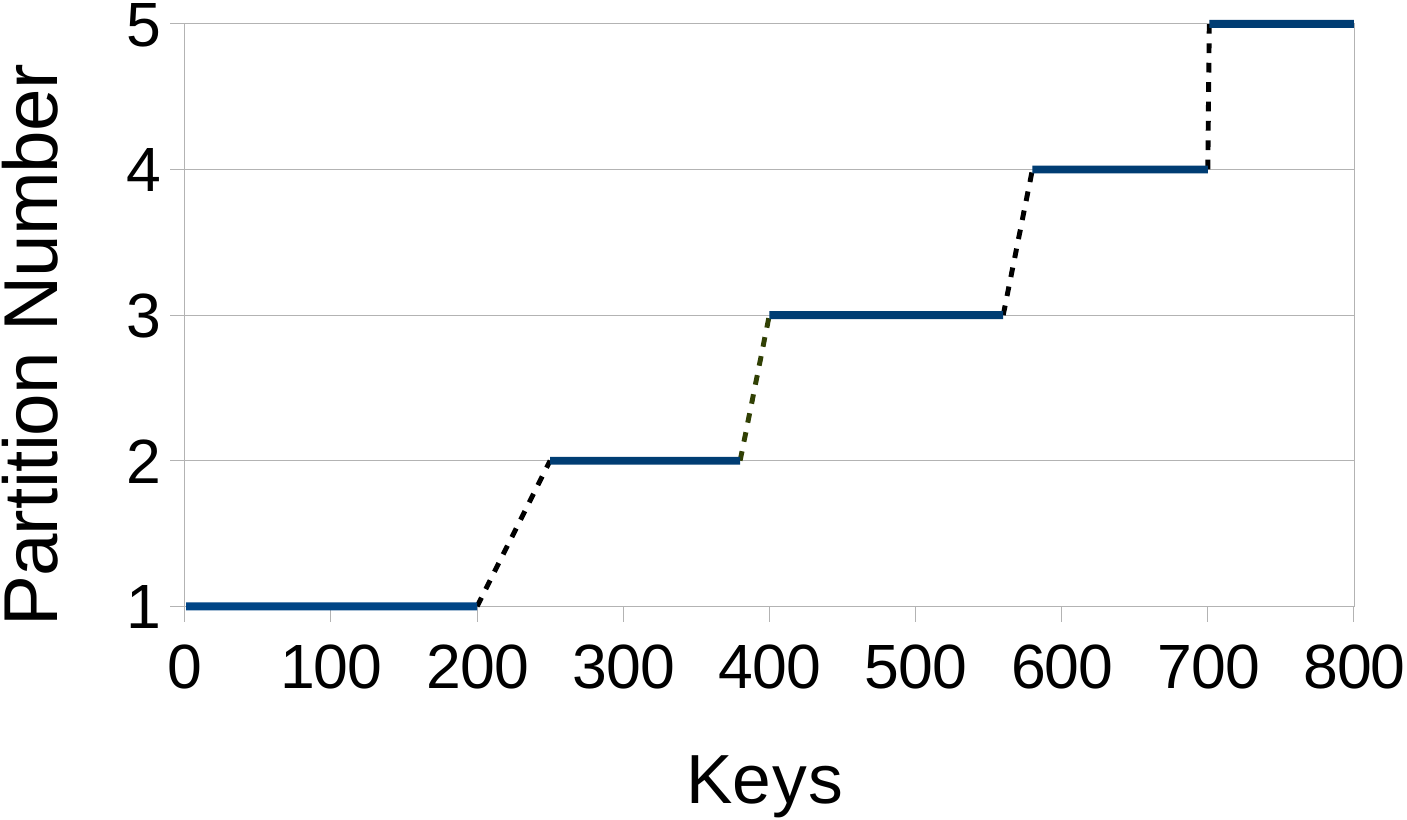}
\captionof{figure}{Keys vs Partition plot}
\label{fig:step_function}
\end{minipage}
\end{table}
  
%
\noindent It can be seen that the plot is a step function \textit{\textbf{f}}, where

\begin{equation} \label{eqn:step_function}
f(key) = \begin{cases} 
      1 & 1 \leq key \leq 200 \\
      2 & 250 \leq key \leq 380 \\
      3 & 400 \leq key \leq 560 \\
      4 & 580 \leq key \leq 700 \\
      5 & 701 \leq key \leq 800 \\
   \end{cases}
\end{equation}

\noindent which can equivalently be represented by summing several \textit{Partition Functions}, which constitutes the \textit{Learned  Index  Function} $L_{customer}$ on the \textbf{customer} table as  
\begin{multline*}
L_{customer}(key) = P_{1,200,1}(key) + P_{250,380,2}(key) \\+ P_{400,560,3}(key) + P_{580,700,4}(key) + P_{701,800,5}(key)
\end{multline*}

\noindent where, \textit{a, b} and \textit{c} of the \textit{Partition Function P} represent the begin and end keys and the corresponding partition id. 

Next we explain how our learned index can effectively be used for join queries. Let \textit{A} and \textit{B} be two relations, which need to be  joined on $col_A$ and $col_B$ of \textit{A} and \textit{B} respectively. Without loss of generality, we assume $col_A$ is sorted and a \textit{Learned  Index  Function} ($L_A$) is generated on this column. Since DaskDB internally uses Dask APIs, before joining the relations, they are converted to Dask dataframes. Each dask dataframes consists of several partitions. For each tuple \textit{\textbf{t}} of a partition of \textit{B}, $L_A(t[col_B])$ is calculated (which returns the partition number of \textit{A} to which $t[col_B]$ belongs) and is appended to \textit{\textbf{t}} as a new \lq Partition' column. This process is parallely executed for each partition of \textit{B}. Then for each partition number \textit{\textbf{i}} of \textit{A}, all the tuples \textit{\textbf{t}} of \textit{B} are selected such that $t['Partition'] = i$ and are hence joined with the $i^{th}$ partition of A. The processes of partition identification, partition selection and the partition-wise join are individually executed in parallel. For joining the dataframes, we developed a variant of the \code{merge} API of  
Pandas package.

%% file: evaluation.tex
\section{Evaluation}
\label{sec:Dask_DBEvaluation}
In this section, 
we present the experimental setup, TPC-H benchmark results and a custom UDF benchmark results.

\subsection{Experimental Setup}

DaskDB was implemented in Python by extending Dask. The SQLParser  of DaskDB utilizes the tool sql-metadata \cite{url:sql-metadata-url}. The QueryPlanner of DaskDB extends  Raco \cite{Wang2017Myria}.  We ran experiments on a cluster of 8 machines each having 16 GB memory and 8 Intel(R) Xeon(R) CPUs running at 3.00 GHz and each machine ran Ubuntu 16.04 OS.

We evaluated DaskDB against two systems that support both SQL query execution and data analytics: PySpark and Hive/Hivemall (\textit{henceforth referred to as Hivemall}).  HDFS was used to store the datasets for each system.
The software versions of Python, PySpark and Hive were 3.7.6,  3.0.1 and 2.1.0 respectively. PySpark and Hivemall were allocated maximum resources available (i.e. cores and memory).

\begin{figure*}[ht]
\centering
  \begin{subfigure}[b]{0.2\textwidth}
	   \centering
	   \includegraphics[width=1.2\textwidth]{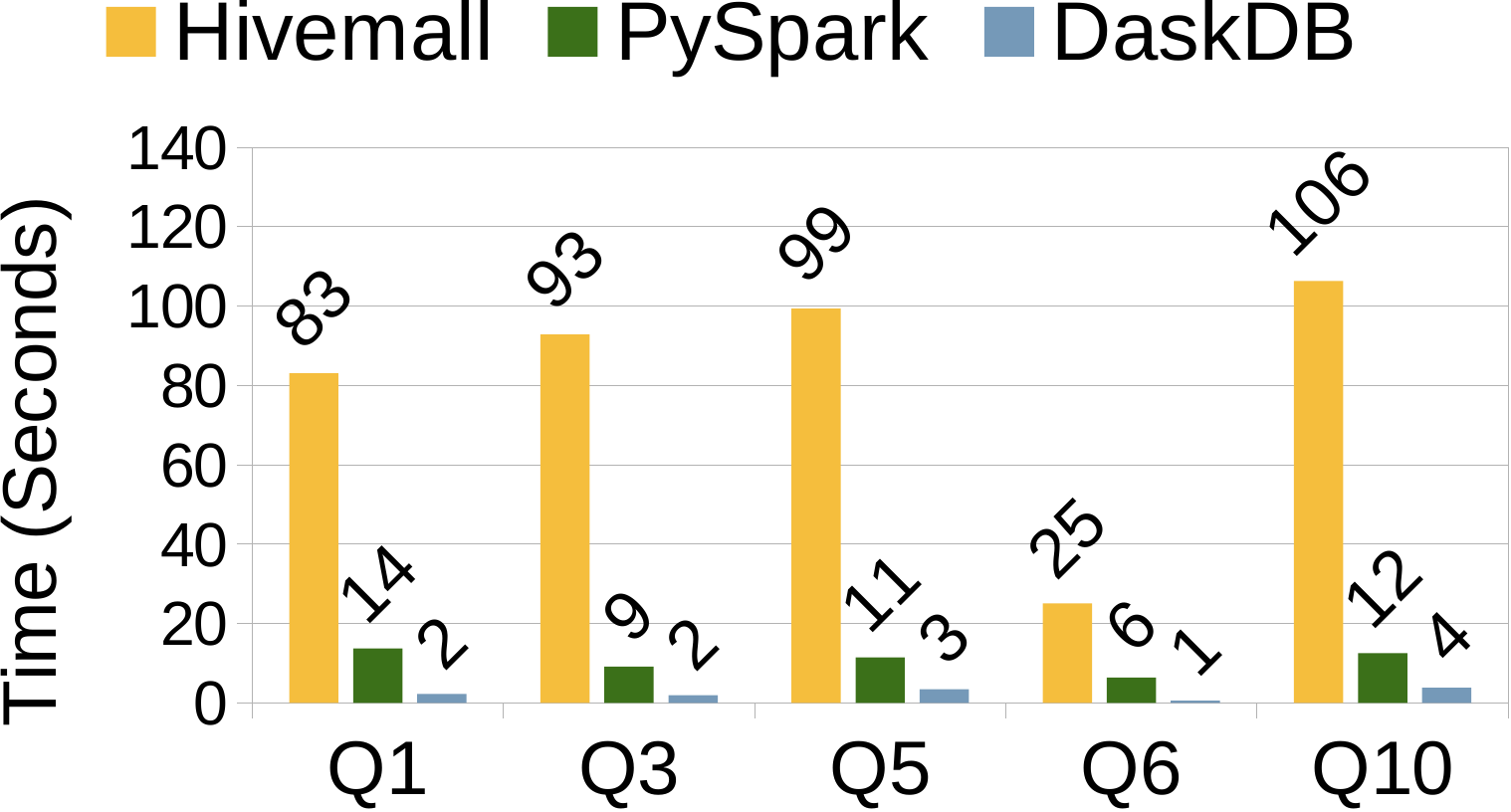}
	    \label{subfig:tpch1}
	    \caption{Scale Factor 1 (GB)}
	\end{subfigure}    
 \hfill
  \begin{subfigure}[b]{0.2\textwidth}
	   \centering
	   \includegraphics[width=1.2\textwidth]{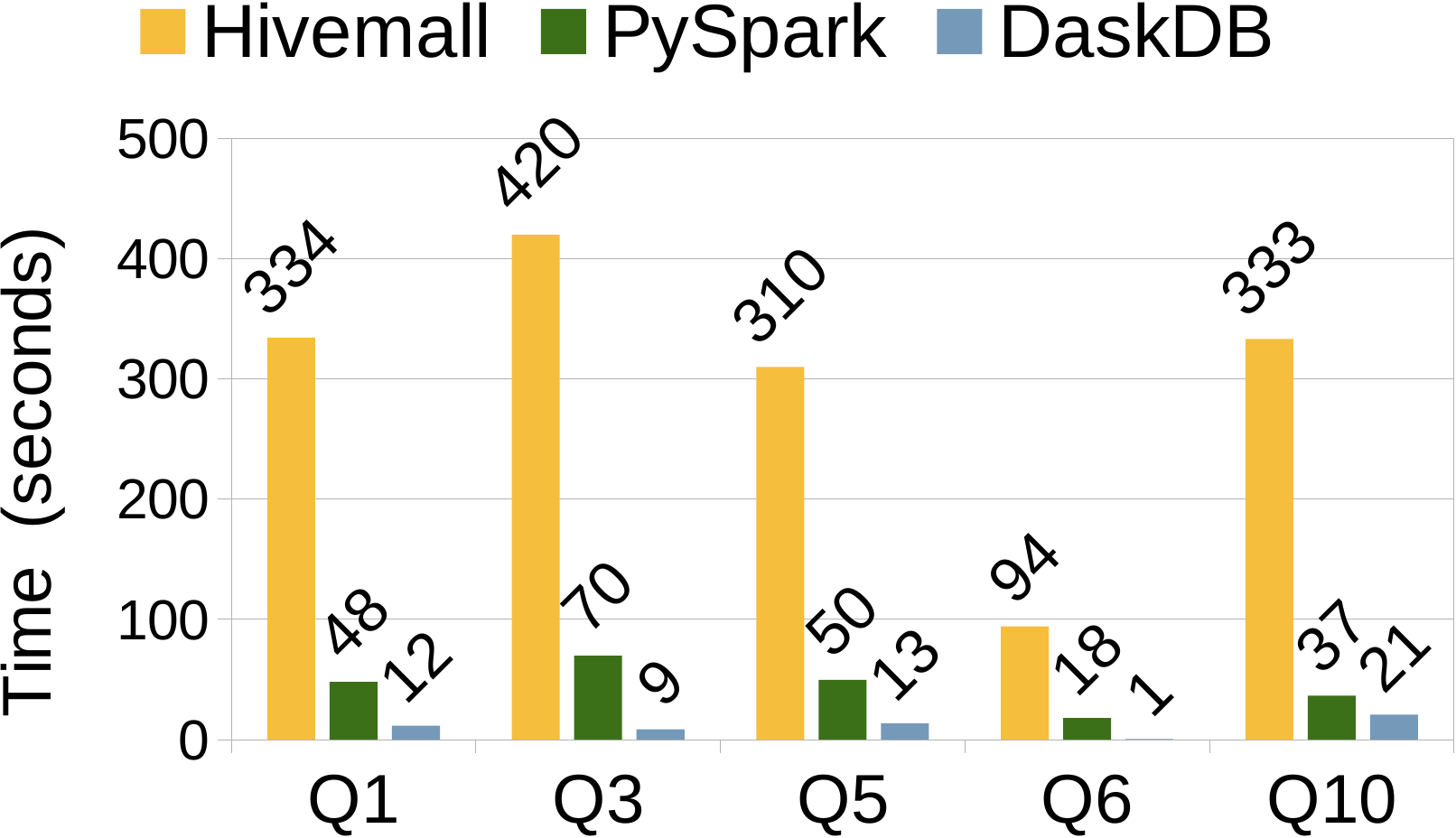}
	    \label{subfig:tpch5}
	    \caption{Scale Factor 5 (GB)}
	\end{subfigure}    
 \hfill
  \begin{subfigure}[b]{0.2\textwidth}
	   \centering
	   \includegraphics[width=1.2\textwidth]{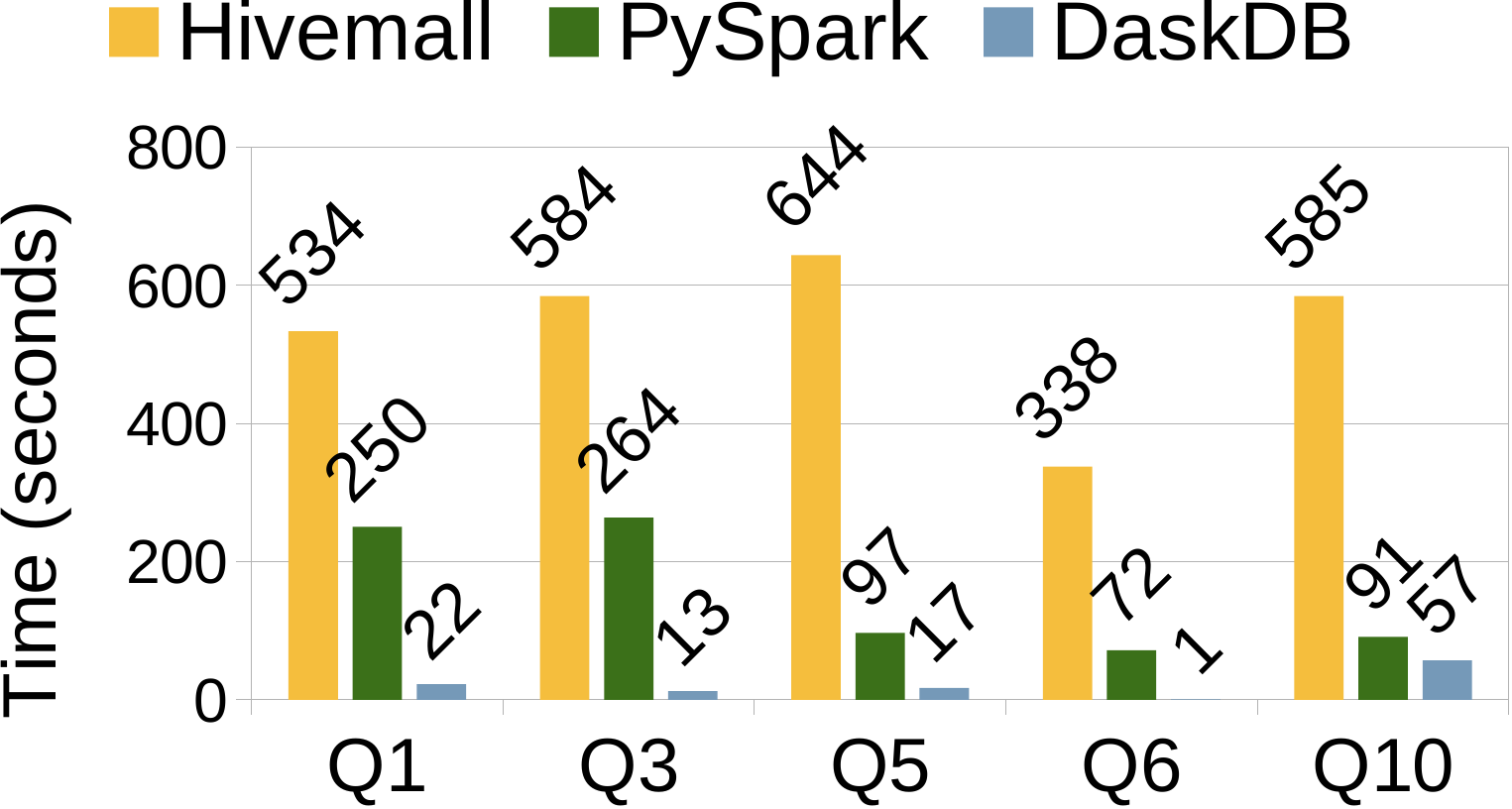}
	    \label{subfig:tpch10}
	    \caption{Scale Factor 10 (GB)}
	\end{subfigure}    
 \hfill
  \begin{subfigure}[b]{0.2\textwidth}
	   \centering
	   \includegraphics[width=1.2\textwidth]{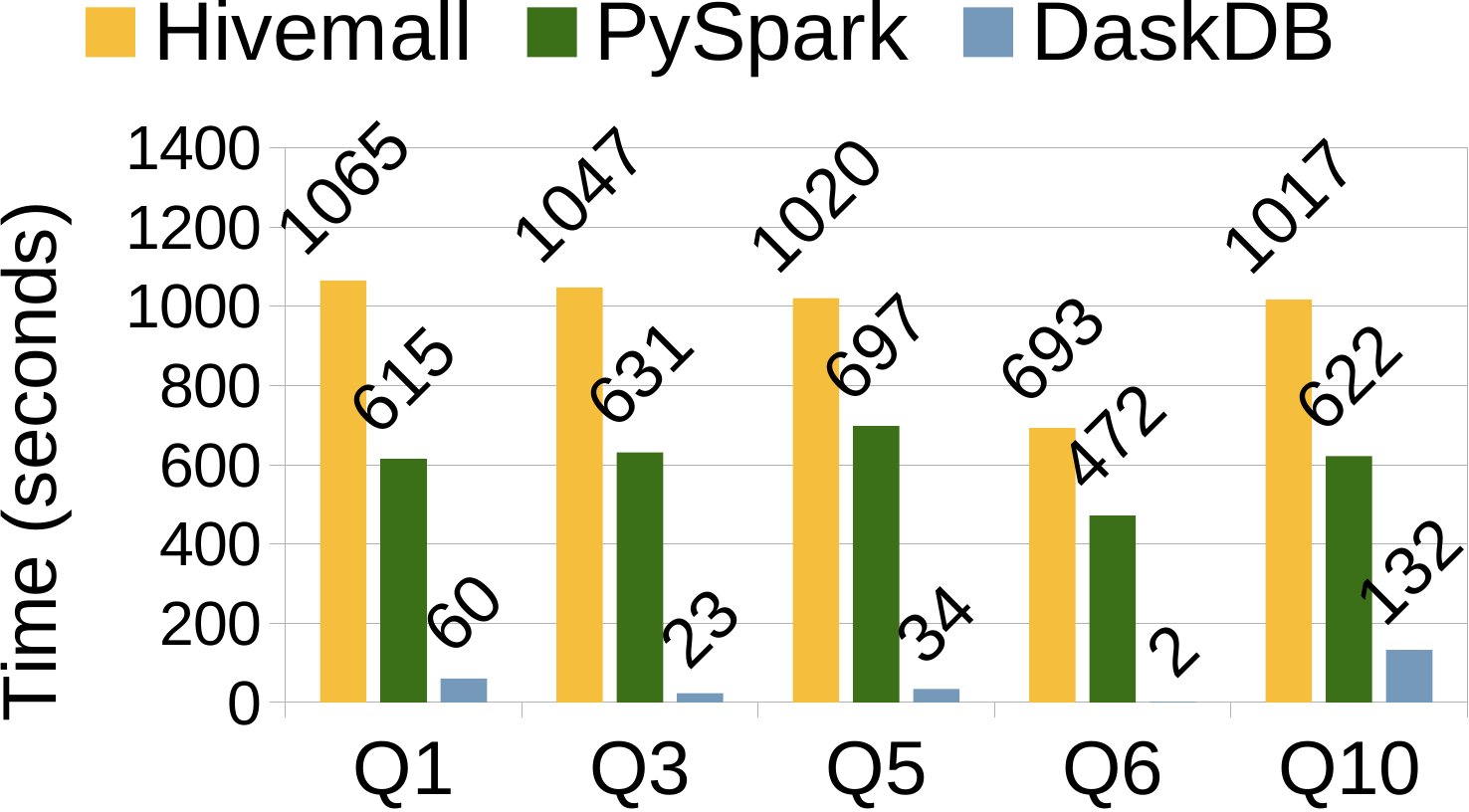}
	    \label{subfig:tpch20}
	    \caption{Scale Factor 20 (GB)}
	\end{subfigure}    
\caption{Execution times - TPC-H benchmark queries}
\label{fig:tpch}
\end{figure*}

\begin{figure*}
  \subfloat[UDF on SF 1]{
	\begin{minipage}[b]{0.2\textwidth}
	   \centering
	   \includegraphics[width=1.2\textwidth]{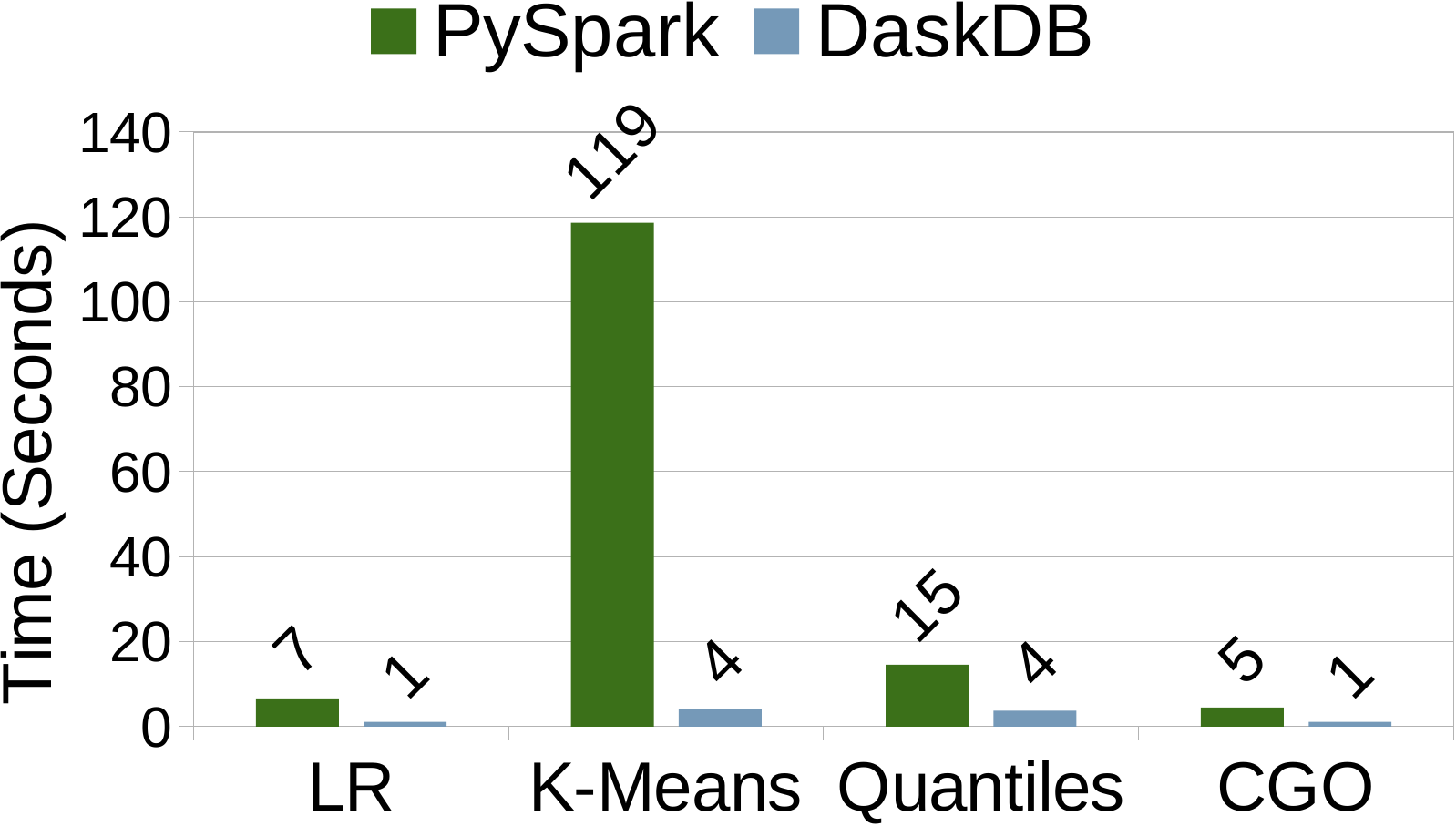}
	\end{minipage}
	\label{subfig:udf1}}
 \hfill
  \subfloat[UDF on SF 5]{
	\begin{minipage}[b]{0.2\textwidth}
	   \centering
	   \includegraphics[width=1.2\textwidth]{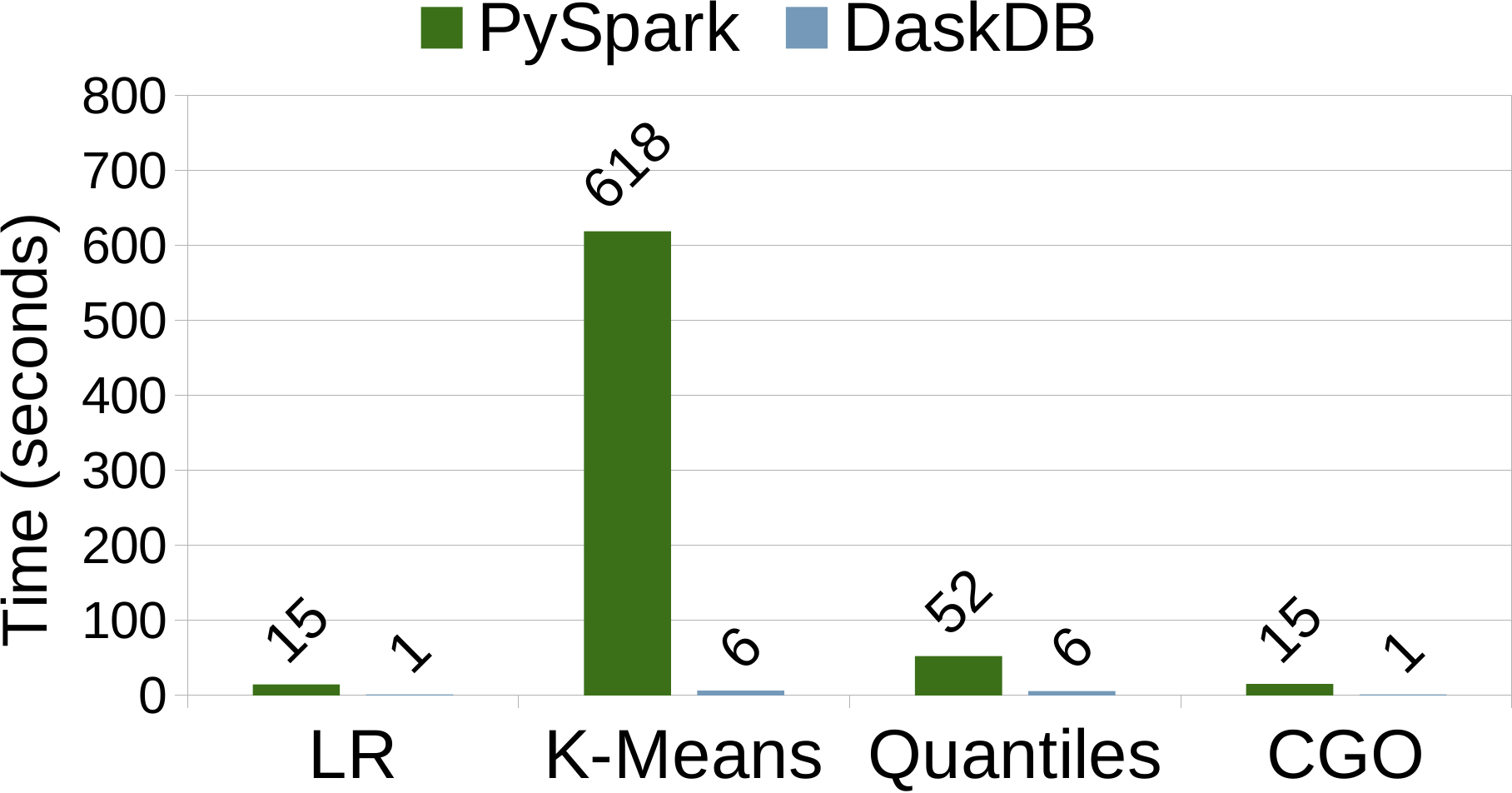}
	\end{minipage}
	\label{subfig:udf5}}
\hfill	
  \subfloat[UDF on SF 10]{
	\begin{minipage}[b]{0.2\textwidth}
	   \centering
	   \includegraphics[width=1.2\textwidth]{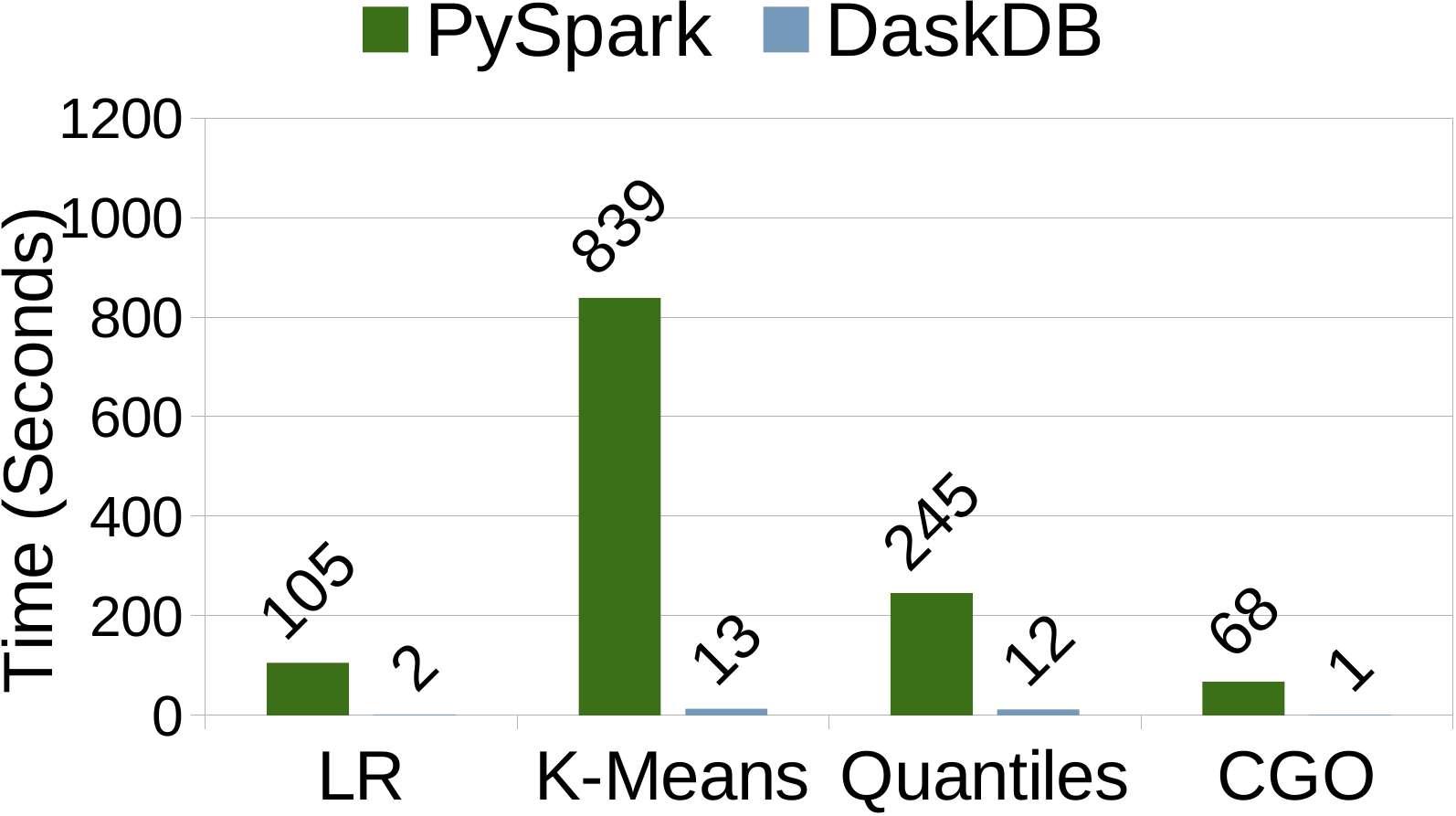}
	\end{minipage}
	\label{subfig:udf10}}
 \hfill	
  \subfloat[UDF on SF 20]{
	\begin{minipage}[b]{0.2\textwidth}
	   \centering
	   \includegraphics[width=1.2\textwidth]{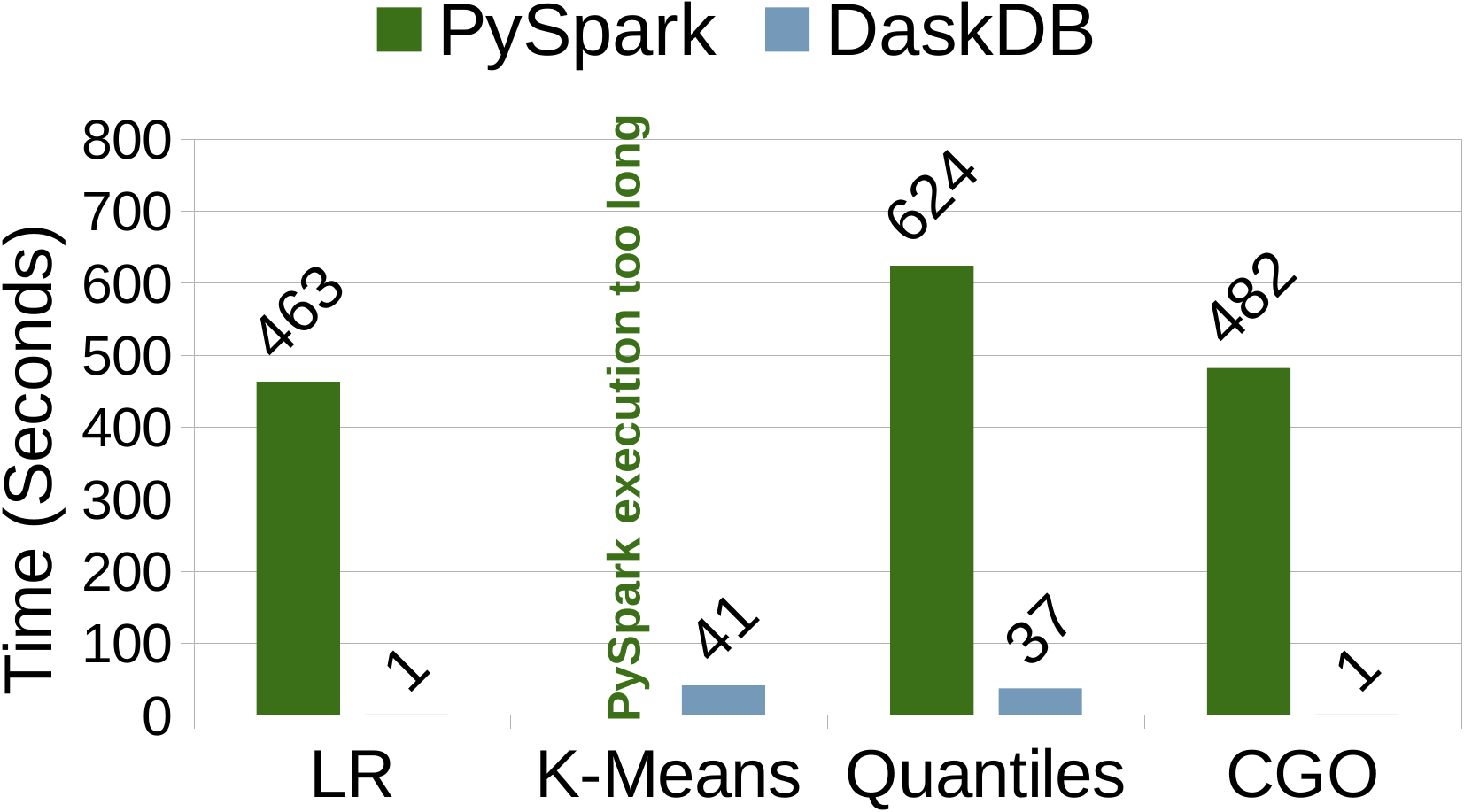}
	\end{minipage}
	\label{subfig:udf20}}
\caption{Execution times - SQL queries with UDFs}
\label{fig:UDF_on_SF}
\vspace{-8pt}
\end{figure*}

\subsection{TPC-H Benchmark Evaluation Results}

We evaluated the systems with several queries from TPC-H decision support  benchmark ~\cite{url:TPCH}.  We used 4 scale factors (SF): 1, 5, 10 and 20, where SF 1 indicates roughly 1 GB.

We executed 5 queries from TPC-H benchmark and the results are plotted in  Figure \ref{fig:tpch}.
As can be seen, DaskDB outperforms PySpark and Hivemall on all queries for all the scale factors. Hivemall performs worse than both DaskDB and PySpark in all cases.
 However, in general higher  the SF,  larger the performance gap between them. For instance, with Q10, DaskDB is 3.5$\times$ faster than PySpark at SF 1 and 4.7$\times$ faster than PySpark at SF 20. DaskDB achieves a speedup of 182$\times$ with Q5 at SF 20.  The superior performance of DaskDB can be credited to its efficient data distribution,  join implementation using distributed learned index and selective data persistence.


\subsection{UDF Benchmark Evaluation Results}
We developed a custom UDF benchmark, which consists of four  machine learning tasks with UDFs:   \textbf{LR (Linear Regression), K-Means (K-Means Clustering), Quantiles (Quantiles Estimation)} and \textbf{CGO (Conjugate Gradient Optimization)}. 
They were developed using the available machine learning packages in Python. For each of the machine learning tasks, 
 SQL queries in Table \ref{table:queries_with_UDF} were executed.
The UDFs were written to perform the same task in both the systems by importing the same Python packages.
For this evaluation, the performance of DaskDB was compared with that of PySpark, whereas Hivemall results are skipped due to poor performance.   

The results are plotted in Figure \ref{fig:UDF_on_SF}. Similar to the TPC-H benchmark results, DaskDB outperforms PySpark here as well for all the machine learning tasks for all scale factors. Among all the tasks K-Means performs worst in PySpark. With respect to K-Means, DaskDB performs 28.5$\times$ faster than PySpark at SF 1 and 64$\times$ faster at SF 10. For SF 20, PySpark took too long to perform K-Means and hence could not be measured, whereas DaskDB took only 41s approximately. For Quantiles, DaskDB was 4$\times$ and 16.6$\times$ faster than PySpark for SF 1 and SF 20 respectively. 
These results also show that larger the SF, the better DaskDB performs compared to PySpark.


\sd{
}

%% file: conclusion.tex
\section{Conclusion}
\label{sec:conclusion}

We presented DaskDB, a scalable data science system.
It brings \textit{in situ} SQL querying to a data science platform in a way
that supports high usability, performance, scalability and built-in capabilities. Moreover, DaskDB also has the ability to incorporate any UDF into the input SQL query, where the UDF could invoke any Python library call. Furthermore, we introduce a novel distributed learned index that accelerates join/merge operation. 
We evaluated 
DaskDB against two state-of-the-art systems, PySpark and Hive/Hivemall, using TPC-H benchmark and a custom UDF benchmark. We show that DaskDB significantly outperforms these  systems.